\documentclass[twocolumn]{aastex63}

\submitjournal{ApJ}

\shorttitle{Variability in Molecular Ion Emission}
\shortauthors{Waggoner et al.}

\usepackage{listings}
\usepackage{hyperref}
\usepackage[caption=false]{subfig}

\begin{document}

\title{MAPS: Constraining Serendipitous Time Variability in Protoplanetary Disk Molecular Ion Emission}

\correspondingauthor{Abygail R. Waggoner}
\email{arw6qz@virginia.edu}

\author[0000-0002-1566-389X]{Abygail R. Waggoner}
\affiliation{University of Virginia, Charlottesville, VA 22904, USA}
\author[0000-0003-2076-8001]{L. Ilsedore Cleeves}
\affiliation{University of Virginia, Charlottesville, VA 22904, USA}
\author[0000-0002-8932-1219]{Ryan A. Loomis}
\affiliation{National Radio Astronomy Observatory, 520 Edgemont Rd., Charlottesville, VA 22903, USA}
\author[0000-0003-3283-6884 ]{Yuri Aikawa}
\affiliation{Department of Astronomy, Graduate School of Science, The University of Tokyo, Tokyo 113-0033, Japan}
\author[0000-0001-7258-770X]{Jaehan Bae}
\affiliation{Department of Astronomy, University of Florida, Gainesville, FL 32611, USA}
\author[0000-0002-8716-0482]{Jennifer B. Bergner}
\affiliation{UC Berkeley Department of Chemistry, Berkeley CA 94720}
\author[0000-0003-2014-2121]{Alice S. Booth}
\affiliation{Leiden Observatory, Leiden University, 2300 RA Leiden, the Netherlands}
\affiliation{School of Physics and Astronomy, University of Leeds, Leeds, UK, LS2 9JT}
\author[0000-0002-0150-0125]{Jenny K. Calahan}
\affiliation{Department of Astronomy, University of Michigan, 323 West Hall, 1085 South University Avenue, Ann Arbor, MI 48109, USA}
\author[0000-0002-2700-9676 ]{Gianni Cataldi}
\affil{National Astronomical Observatory of Japan, Osawa 2-21-1, Mitaka, Tokyo 181-8588, Japan}
\affil{Department of Astronomy, Graduate School of Science, The University of Tokyo, Tokyo 113-0033, Japan}
\author[0000-0003-1413-1776 ]{Charles J. Law}
\affiliation{Center for Astrophysics | Harvard \& Smithsonian, 60 Garden St., Cambridge, MA 02138, USA}
\author[0000-0003-1837-3772 ]{Romane Le Gal}
\affiliation{University Grenoble Alpes, CNRS, IPAG, F-38000 Grenoble, France}
\affiliation{RAM, 300 rue de la piscine, F-38406 Saint-Martin d’H\`{e}res, France}
\author[0000-0002-7607-719X]{Feng Long}
\altaffiliation{NASA Hubble Fellowship Program Sagan Fellow}
\affiliation{Lunar and Planetary Laboratory, University of Arizona, Tucson, AZ 85721, USA}
\author[0000-0001-8798-1347]{Karin I. \"{O}berg}
\affiliation{Center for Astrophysics | Harvard \& Smithsonian, 60 Garden St., Cambridge, MA 02138, USA}
\author[0000-0003-1534-5186 ]{Richard Teague}
\affiliation{Department of Earth, Atmospheric, and Planetary Sciences, Massachusetts Institute of Technology, Cambridge, MA 02139, USA}
\affiliation{Center for Astrophysics | Harvard \& Smithsonian, 60 Garden St., Cambridge, MA 02138, USA}
\author[0000-0003-1526-7587 ]{David J. Wilner}
\affiliation{Center for Astrophysics | Harvard \& Smithsonian, 60 Garden St., Cambridge, MA 02138, USA}

\begin{abstract} 

Theoretical models and observations suggest that the abundances of molecular ions in protoplanetary disks should be highly sensitive to the variable ionization conditions set by the young central star. 
We present a search for temporal flux variability of HCO$^+$ J$=1-0$, which was observed as a part of the Molecules with ALMA at Planet-forming Scales (MAPS) ALMA Large Program. 
We split out and imaged the line and continuum data for each individual day the five sources were observed 
(HD~163296, AS~209, GM~Aur, MWC~480, and IM~Lup, with between 3 to 6 unique visits per source).  
Significant enhancement ($>3\sigma$) was not observed, but we find variations in the spectral profiles in all five disks.
Variations in AS~209, GM~Aur, {and} HD~163296 are tentatively attributed to variations in HCO$^+$ flux, while variations in IM~Lup and MWC~480 are most likely introduced by differences in the \textit{uv} coverage, which impact the amount of recovered flux during imaging.
The tentative detections and low degree of variability are consistent with expectations of X-ray flare driven HCO$^+$ variability, which requires relatively large flares to enhance the HCO$^+$ rotational emission at significant ($>20\%$) levels. These findings also demonstrate the need for dedicated monitoring campaigns with high signal to noise ratios to fully characterize X-ray flare driven chemistry. 

\end{abstract}

\keywords{protoplanetary disk, astrochemistry, radio astronomy}

\section{Introduction} 
\label{sec:intro}

Molecular abundances within protoplanetary disks have traditionally been expected to evolve over tens to hundreds of thousands of years or longer \citep[][and references therein]{henning2013,oberg2021}. 
However, young pre-main sequence stars at the center of disks are active in the X-ray regime on timescales of days to weeks \citep[e.g.][]{getman2005,wolk2005,getman2022a,getman2022c}.
Due to an unstable dynamo, young stars regularly undergo magnetic reconnection events which result in a larger burst of X-ray photons commonly known as an X-ray flare \citep[][and references therein]{guedel2004}, which can temporarily increase the X-ray flux and disk ionization rates \citep{Ilgner2006flares}. Flares can range in strength, increasing the flux by a factors of a few to several hundred times the baseline flux \citep{getman2005, getman2021a}

Flare-driven variable ionization rates drive variability in chemical species in disks.
This variation largely stems from time variability in the ionization of H$_2$ and He, which play a major role in driving cold molecular chemistry \citep{maloney1996}. 
For example, a single strong flare (i.e., 100 times stronger than the baseline luminosity) can temporarily increase the abundance of H$_2^+$ by up to a factor of $\sim 70$. H$_2^+$ then forms H$_3^+$, which enhances the formation of gas-phase species, such as H$_2$O, near the disk surface \citep[$<100$~au from the central star][]{waggoner2019}.
Gas-phase cations are especially sensitive to flares, where some species, such as HCO$^+$ and N$_2$H$^+$, can be increased by several orders of magnitude for days or weeks depending on the strength of the flare. 
For example, \citet{waggoner2022} found that a single strong flare can increase the disk integrated abundance of HCO$^+$ and N$_2$H$^+$ by $\sim 4$ and $\sim 3$, respectively. 
Additionally, the cumulative effect of thousands of flares aids in the advancement of chemical complexity over the course of hundreds of years by marginally ($\sim 1\%$ change) increasing the production of carbon chains and organosulphides \citep{waggoner2022}. 

While this study is motivated specifically by flare-driven variable ionization rates, it is important to note that other processes can drive variability. For example, accretion outbursts are known to increase UV flux and disk temperature. Observations and models indicate that outbursts evaporate ices and increase gas density, which in turn drives chemical reactions \citep[e.g.,][]{molyarova2018,kospal2021,leemker2021,ruizrodriguez2022,fischer2022,tobin2023}.

Significant enhancement in the flux of a gas-phase cation has been reported previously by \citet{cleeves2017} in observations of the H$^{13}$CO$^+$ J=$3-2$ line with the Atacama Large Millimeter/submillimeter Array (ALMA).
Cleeves et. al observed H$^{13}$CO$^+$ J=$3-2$ in the IM~Lup protoplanetary disk on three separate days: July 2014, January 2015, and May 2015. In July 2014 and January 2015, the H$^{13}$CO$^+$ flux was effectively constant, but in May 2015 the flux doubled (an increase by $28 \sigma$).
They concluded that enhancement was not caused by an increase in temperature, as the continuum emission remained constant, thus ruling out an outburst or similar type event. 
IM~Lup, like most young stars, is known to be X-ray variable \citep{cleeves2017}.
Indeed, the most likely source of H$^{13}$CO$^+$ enhancement was due to an X-ray flaring event.

In fact, the detection in IM~Lup was serendipitous and is currently the only source with reported variability.
{\citet{espaillat2022} searched} for variations in CO millimeter emission in GM~Aur; however no significant variability was seen. This result was somewhat expected, since chemical models {in \citet{espaillat2022} and} in \citet{waggoner2022} indicate that the robust CO molecule is not sensitive to variable X-ray and UV ionization rates.

This work seeks evidence of flare driven chemistry by searching for variability of the HCO$^+$ J=$1-0$ line in the five disks, HD~163296, AS~209, GM~Aur, MWC~480, and IM~Lup, observed as a part of Molecules with ALMA at Planet-forming Scales (MAPS) ALMA Large Program \citep{maps1}, which fortuitously was observed at multiple dates spanning about a year. 
Section \ref{sec:observations} describes the observational set up, sources, and line selection. 
Section \ref{sec:methods} describes the {\ttfamily CLEAN}ing strategy, methodology, and error analysis. 
The results are presented in Section \ref{sec:results}, where the degree of variability is constrained. 
In Section \ref{sec:discussion}, we discuss the presence/absence of variability, along with possible connections to X-ray flaring events and/or related variable ionization rates. 
Section \ref{sec:conclusions} provides a summary of this work along with concluding remarks. 

\begin{table*}[]
    \centering
    \begin{tabular}{c|c|c|c|c|c|c|c}
         Source     &  R.A.               & Decl.               & Sys Vel              & Num of    & Min Baseline      & Beam      & BPA\\
                    &    (J2000)          &                     & (LSRK, km~s$^{-1}$)  & Obs.      & (m)               & ''        &    \\
                    \hline
         IM~Lup     & $15:56:09.186780$   & $-37.56.06.58091$   & $4.5^{(6)}$           & $3$         & $60$            & $2.0$     & 63.8   \\
         GM~Aur     & $04:55:10.981558$   & $+30.21.58.87742$   & $5.6^{(4)}$          & $5$         & $55$            & $2.0$     & 8.6  \\
         AS~209     & $16:49:15.293780$   & $-14.22.09.09404$   & $4.6^{(3)}$           & $5$         & $50$            & $2.0$     & 62.4  \\
         HD~163296  & $17:56:21.277330$   & $-21.57.22.63945$   & $5.8^{(1,2)}$        & $6$         & $70$            & $2.0$     & 76.3  \\
         MWC~480    & $04:58:46.274800$   & $+29.50.36.47709$   & $5.1^{(2,5)}$        & $5$         & $50$            & $2.2$     & 16.7  \\
    \end{tabular}
    \caption{The right ascension,  declination, systemic velocity, number of unique observations days, minimum baseline used for imaging, and beam size for each of the five MAPS disks. For each disk, the beam was tapered and smoothed to be circular and the same across all epochs. The minimum baseline is the smallest baseline coverage used in the imaging process. 
    (1) \citet{teague2019}
    (2) \citet{teague2021}
    (3) \citet{huang2017}
    (4) \citet{huang2020}
    (5) \citet{pietu2007}
    (6) \citet{pinte2018a}
    }
    \label{tab:diskparams}
\end{table*}

\begin{table}[]
    \centering
    \begin{tabular}{cccc} 
Source	    & Date	            & \textit{uv}	& time \\
	        &		            &    (m)        & 	   (s)          \\
	        \hline
IM~Lup	    & 29 Oct 2018	    & 13.5-1258.6	& 2470.90        \\
	        & 20 Aug 2019	    & 35.7-3188.1	& 3751.49       \\
	        & 21 Aug 2019	    & 39.2-3506.6	& 3698.02       \\
         	 \hline
GM~Aur	    & 13 Dec 2018	    & 12.7-667.3	& 2372.98       \\
	        & 15 Dec 2018	    & 12.0-687.8	& 2345.52       \\
	        & 31 Aug 2019	    & 22.7-3143.7	& 3976.56       \\
	        & 02 Sep 2019	    & 24.4-3141.3	& 8548.70        \\
	        & 04 Sep 2019	    & 25.7-2904.0	& 4218.82       \\
	   \hline 
AS~209	    & 26 Oct 2018	    & 14.3-1394.8	& 2520.43       \\
	        & 23 Aug 2019	    & 39.9-3396.5	& 3660.77       \\
	        & 24 Aug 2019	    & 31.6-3330.3	& 3612.77       \\
	        & 03 Sep 2019	    & 70.5-3118.4	& 3666.29       \\
	        & 04 Sep 2019	    & 27.7-3627.5	& 3641.57       \\
	  \hline  
HD~163296	& 22 Oct 2018	    & 12.7-1347.6	& 2511.17     \\
	        & 23 Aug 2019	    & 56.7-2742.3  	& 3503.42         \\
	        & 24 Aug 2019	    & 30.3-3335.1	& 3499.78       \\
	        & 25 Aug 2019	    & 39.1-3389.0	& 3519.79       \\
	        & 04 Sep 2019	    & 34.6-3071.5	& 3508.90        \\
	        & 05 Sep 2019	    & 73.0-3637.1	& 3527.23       \\
	   \hline   
MWC~480	    & 13 Dec 2018	    & 12.0-671.2	& 2974.51       \\
	        & 15 Dec 2018	    & 12.1-691.7	& 2931.70        \\
	        & 31 Aug 2019	    & 23.0-3143.8	& 4141.15       \\
	        & 02 Sep 2019	    & 24.7-3142.6	& 8573.71       \\
	        & 04 Sep 2019	    & 26.1-2884.5	& 4245.26       \\                
    \end{tabular}
    \caption{The baseline coverage (\textit{uv}) and total observation time spent on source for each observation day for each disk. Each unique observation day was a single ALMA execution block, except 02/Sep/2019 in MWC~480 and GM~Aur, which contain two execution blocks.}
    \label{tab:diskdates}
\end{table}

\section{Observations}\label{sec:observations}

The data used in this work were taken as a part of the MAPS ALMA Large Program \citep[2018.1.01055.L, ][]{maps1}.
MAPS observations included five sources: the IM~Lup, GM~Aur, AS~209, HD~163296, and MWC~480 protoplanetary disks. Table \ref{tab:diskparams} includes the coordinates, systemic velocity, and additional observation details for each disk. 
MWC~480 and HD~163296 are both Herbig Ae systems, 
while the other three are T Tauri systems.
These five disks were selected because they are bright and chemically rich, thus allowing for more in-depth observations \citep[][Section 2.1 for more details]{maps1}.

HCO$^+$ J=$1-0$ observations were carried out between October 2018 and September 2019, where short-baseline data were collected in October  and December 2018 and long-baseline data were taken in August and September 2019. Table \ref{tab:diskdates} provides the baseline coverage and integrated observation time for each observation of each disk. 
A detailed description of data collection and the data reduction analysis is provided in \citet{maps1} and \citet{maps2}, respectively. 

For the most part, there was only a single execution per observation day, with the exception of MWC~480 and GM~Aur. Both of these sources were observed twice on 02/Sep/2019. 
A one day separation was found to be the optimal time to split the data based on: 1. timing of the execution blocks, 2. optimizing the signal to noise ratio of the data, and 3. providing the maximum number of distinct observations possible to search for variability.

For this work, we chose to limit the data to HCO$^+$ $1-0$ taken as a part of MAPS. There are previous observations of HCO$^+$ emission lines and its various isotopes in the MAPS sources.
However, by limiting the analysis to MAPS data the analysis is much more homogeneous and allows for a more robust comparison between observations.

\subsection{Line Selection}

We use Band 3 data of HCO$^+$ J=$1-0$ ($89.188525$ GHz) for each of the five MAPS disks. 
HCO$^+$ was found to be the most likely candidate to trace variable disk chemistry based on models used in \citet{waggoner2022}, as gas-phase cations have been shown to be the most susceptible to variable ionization rates. \citet{maps13} found that {the optical depth ($\tau$) of the HCO$^+$ line is $\sim 1$ for} each of the MAPS disks. While a more optically thin line would be more ideal for this study, HCO$^+$ $1-0$ is the only gas-phase cation that is sufficiently bright to allow splitting into individual images for each observation block.

We investigated whether a similar analysis was possible for neutral species observed simultaneously with HCO$^+$; however, the C$_2$H and HCN lines were too weak to split into individual observations. {A search for minor spectral changes was not feasible due to low signal to noise ratios. }

All data used in this work are publicly available directly on the ALMA archive or through the interactive MAPS website\footnote{\href{https://alma-maps.info/}{alma-maps.info}} \citep[See also Section 3.5 in ][]{maps1}. {The line measurement set, line mask, and line plus continuum measurement set can be downloaded directly from the MAPS website.
Continuum masks were hand drawn for each disk.}

\section{Methods}\label{sec:methods}

\subsection{{\ttfamily CLEAN}ing Strategy}\label{sec:clean}

The line and continuum are imaged using the same process starting with the measurement sets (described in Section \ref{sec:observations}) using {\ttfamily CASA} version $6.2$, where the line is flagged from the line plus continuum measurement set to generate the continuum image. 
Line images are produced using the {\ttfamily briggsbwtaper} weighting with a robust value of $0.5$. {\ttfamily briggsbwtaper} is new to {\ttfamily CASA} as of version $6.2$ and handles gridding more accurately than the commonly used {\ttfamily briggs} weighting.
All images are {\ttfamily CLEAN}ed to $3\sigma$, where $\sigma$ is the rms measured in the dirty image. 
The imaging process proceeds as follows: 
\begin{enumerate}
    \item Data for each observation day is split into a new measurement set using the function {\ttfamily split} in {\ttfamily CASA}. 
    Data taken on individual days was not further split because sensitivity would
    drastically decrease. Additionally, a day was found to be sufficient time for a flare to occur between epochs \citep[for flare statistics see][]{wolk2005}. 
    \item A dirty image is generated to determine the {\ttfamily CLEAN}ing threshold and dirty beam size. 
    \item 
    A \textit{uv} taper is estimated from 
    \begin{equation}
        \rm b_{tap} = \left({{b_{dirty}^{-1}-b_{des}^{-1}} \over { (b_{dirty}^{-2}-b_{des}^{-2}  )^{0.5}  }}  \right)^{-1},
    \end{equation}
    where $\rm b_{tap}$ is the \textit{uv} taper, $\rm b_{dirty}$ is the dirty beam, and $\rm b_{des}$ is the desired beam size.
    The taper is unique for each observation execution, but the same taper is used for both line and continuum imaging. The taper is essential to generate a uniform, approximately circular beam size (the `target beam', Table \ref{tab:diskparams}) for each disk. Without a uniform beam, images for individual days could not be directly compared.
    \item 
    The lower resolution images made by applying the taper to the visibilities were {\ttfamily CLEAN}ed using the  MAPS mask pre-smoothed to a larger beam to encompass all of the line emission.
    The MAPS mask is used to select the {\ttfamily CLEAN}ed region; however, the mask is pre-smoothed to the new, larger beam to encompass the full flux of the lower resolution line image. 
    Smoothing is carried out with the {\ttfamily CASA} function {\ttfamily imsmooth}. 
    \item The {\ttfamily CLEAN}ed beam is nearly circular from the \textit{uv} taper, but not perfectly. The final image is then smoothed using {\ttfamily imsmooth} to generate a uniform beam size and shape for each disk (listed in Table \ref{tab:diskparams}).

\end{enumerate}
The {\ttfamily CLEAN}ing scripts are available upon request from A.R.W.

\begin{figure*}
    \centering
    \includegraphics[scale=0.85]{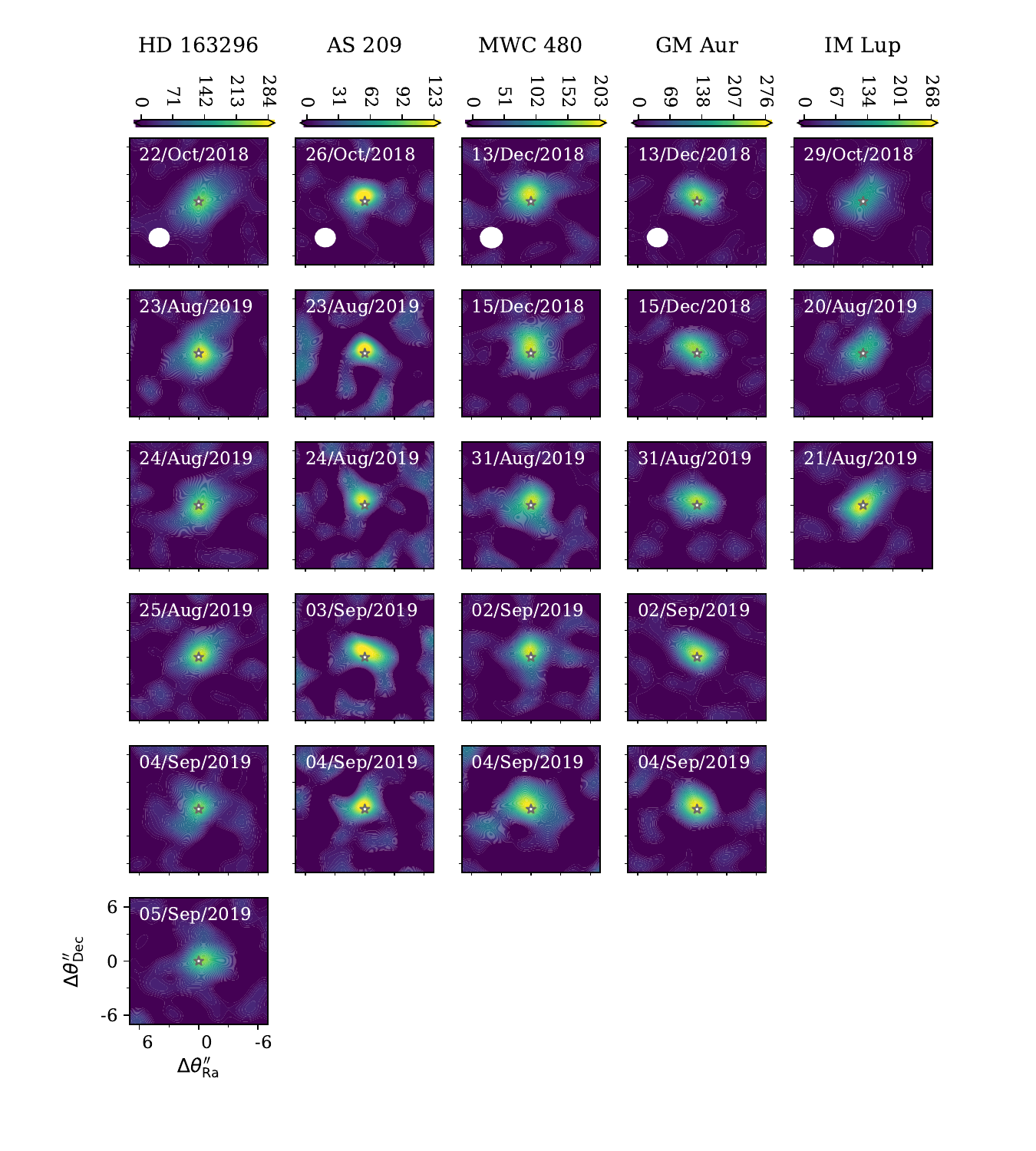}
    \caption{
    The moment zero map of HCO$^+$ $1-0$ for each observation execution for each disk. 
    Each column corresponds to a different disk, while each row represents a separate observation day. The date of observation is shown in the top of each plot. 
    Each disk has a unique own color bar with units in mJy.
    The beam for each disk is shown in white in the bottom left of the top row of images. 
    }    \label{fig:momentzero}
\end{figure*}

{Figure \ref{fig:momentzero}  shows the moment zero maps for all observations, which shows integrated line emission. In this case, the moment zero maps were generated by integrating flux across frequency space across the entire image cube. Maps were generated without use of the mask. 
Figure \ref{fig:continuum}  shows the continuum images for each observation.}

\subsection{Flux and Error Analysis}\label{sec:fluxanderror}

\begin{figure*}
    \centering
    \includegraphics[scale=0.85]{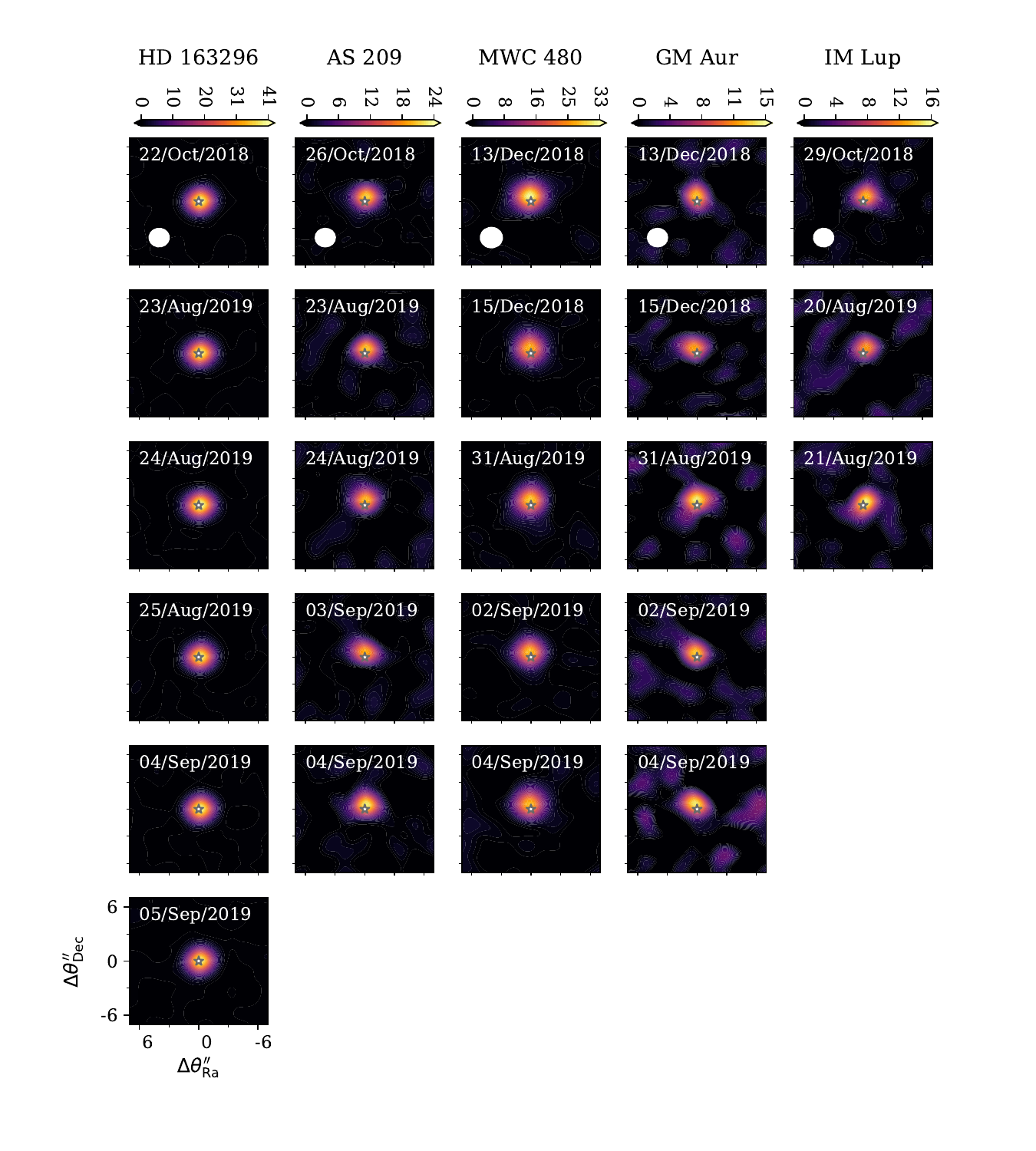}
    \caption{The 90 GHz continuum emission for each observation execution for each disk. 
    Each column corresponds to a different disk, while each row represents a separate observation day. The date of observation is shown in the top of each plot. 
    Each disk has a unique own color bar with units in mJy.
    The beam for each disk is shown in white in the bottom left of the top row of images. 
    }    \label{fig:continuum}
\end{figure*}

The integrated flux for the continuum emission was determined by integrating all emission within the hand-drawn mask {in the continuum image (Figure \ref{fig:continuum})}. 
{
The integrated flux was found by integrating all channels with significant flux}. Channels containing the line were defined as any channel that contain data emission greater than $3\sigma$, where $\sigma$ was the standard deviation of the unmasked data in each channel.

The error for the integrated line flux ($F_{\rm HCO^+}$) was determined as follows.
First, the mask was randomly shifted to $40$ different locations off source. 
The {error of the integrated flux} was then found by taking the standard deviation of the integrated flux within the off source locations. 
The error for the integrated continuum flux ($F_{\rm cont}$) was found in a similar manner. 
This sampling method was used, rather than shifting the mask to line free channels, due to the limited number of channels available in the MAPS products' measurement sets. 
{The flux calibration error is $<20\%$ for all sources, with the assumption that the continuum flux should be constant across all observations 
(Figures \ref{fig:as209_real}-\ref{fig:mwc480_real}). Flux calibration error $\sim 10\%$ or greater is consistent with the ALMA technical handbook \citep{ALMAhandbook}. 
}

To ensure that any changes in HCO$^+$ emission were not due to changes in source brightness or temperature, the line flux to continuum flux ratio ($F_{\rm HCO^+}/F_{\rm cont}$) was used to compare line brightness for individual observation days. 
The continuum is expected to be constant, just as it was observed to be in {for the IM Lup disk in} \citet{cleeves2017}.

\subsection{Baseline Coverage}\label{sec:baselines}

Interferometrically measured flux is inherently sensitive to the baseline coverage of the observation. 
Shorter baselines are known to probe flux better, while being less sensitive to structure, and
longer baselines probe structure well, while being less sensitive to flux \citep[for example see Chapter 5 in ][]{thompson2017}.
The MAPS observations utilized multiple ALMA configurations to optimize the baseline coverage to include both short and long baselines. 
While this method is incredibly useful in generating a complete picture of the disk {across multiple scales}, varying telescope configurations adds an additional challenge to compare flux on different observation days. 

Observations with longer baseline coverage, i.e., fewer short baselines, miss flux on larger scales \citep{thompson2017}. 
{A minimum baseline distance was set for each disk to ensure that long baseline observations do not have an artificially lower flux
than images with thorough short baseline coverage. }
{Even though the most ``flux sensitive'' baselines were removed, there is still sufficient \textit{uv} sampling across all epochs to cross compare them. A range of minimum baseline values was examined, and 
we selected values that preserved as many short baselines as possible consistently across different epochs without cutting out too many short baselines that would introduce imaging artifacts (listed in Table \ref{tab:diskparams}). 
On average there was a $22\% \pm 12$ loss in flux as a result of this process based on comparisons between the clipped data and the full measurement set.
}

To further ensure that any detected variability was not artificially introduced by the varying baseline coverage, the {\ttfamily PYTHON} routine {\ttfamily vis\_sample}\footnote{{\ttfamily vis\_sample} is publicly available under the MIT license at \href{https://github.com/AstroChem/vis\_sample}{https://github.com/AstroChem/vis\_sample} and described in further detail in \citet{loomis2018}} 
was used to create `mock' data sets for each observation of each disk. 
The mock data was generated by first imaging the time integrated images of HCO$^+$ J$=1-0$ (i.e., the complete observation set), as was done by \citet{maps13}. 
The time integrated image is considered to be the `ground truth' to generate the mock data set.
Synthetic ALMA observations were created by running the time integrated image through {\ttfamily vis\_sample} for each of the observation days. This generated a synthetic observation using the same visibility effects as the actual observation.
Mock images were created by imaging the synthetic observations using the same {\ttfamily CLEAN}ing process used for the real data (described above). 
This process helps constrain variability in HCO$^+$ or continuum emission introduced by visibility (i.e., \textit{uv} sampling) effects, but the possibility of visibility effects is still considered in the results. 

\subsection{AS~209}\label{sec:AS209}

\begin{figure*}
    \centering
    \includegraphics[scale=0.5]{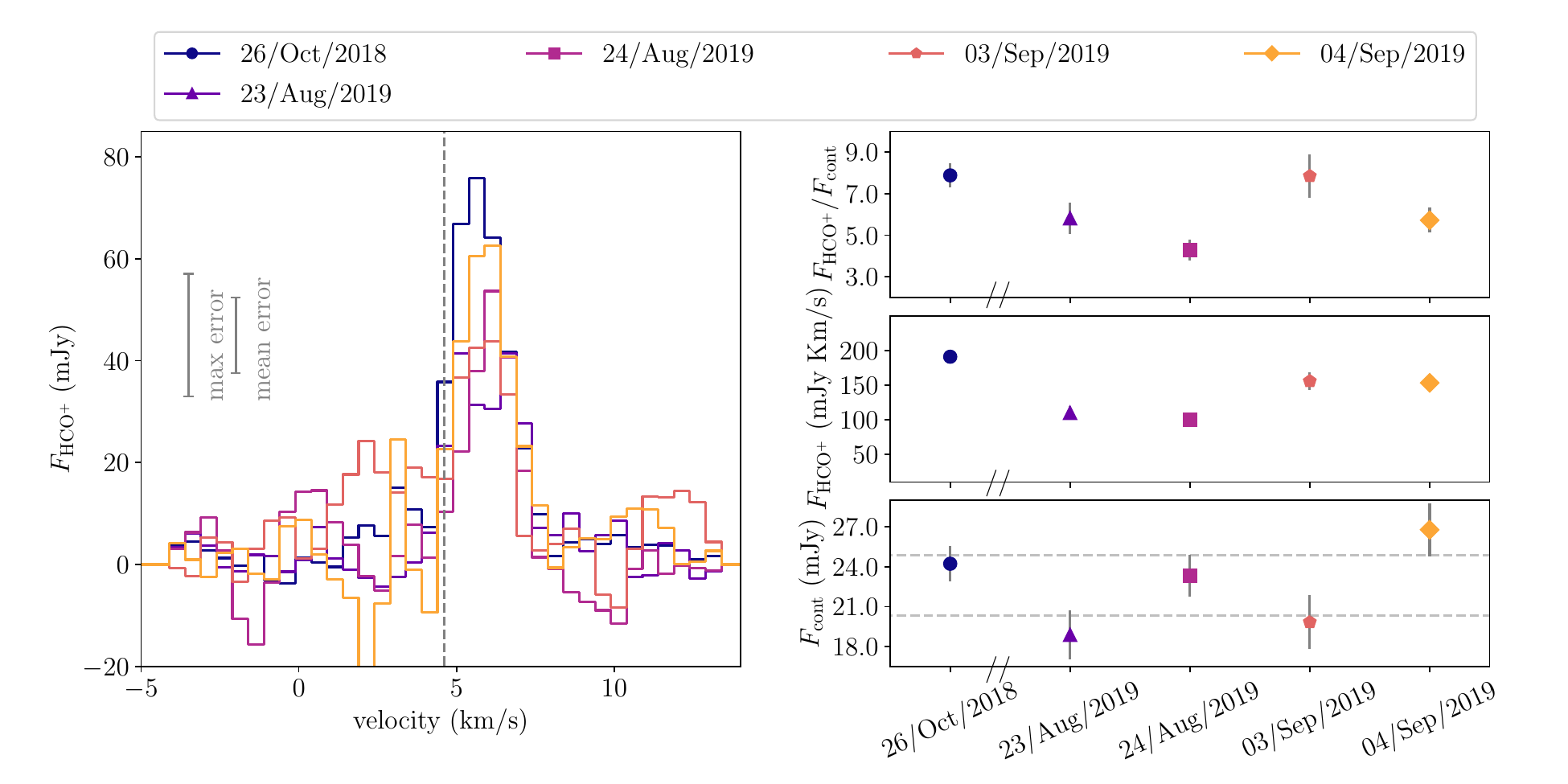}
    \caption{
             \textit{Left}: The HCO$^+$ J=$1-0$ spectra in the AS~209 disk for each unique observation day.  {The grey bars one the left represent the maximum and mean error (standard deviation) for all channels across all observations.} 
             \textit{Right}: The disk integrated continuum flux ($F_{\rm cont}$, bottom), disk integrated HCO$^+$ $1-0$ flux ($F_{\rm HCO^+}$, middle), and normalized line flux with respect to the continuum ($F_{\rm HCO^+}/F_{\rm cont}$, top) for each day HCO$^+$ $1-0$ was observed in AS~209. {Units for $F_{\rm HCO^+}/F_{\rm cont}$ are mJy~km~s$^{-1}$~mJy$^{-1}$.} Error bars indicate $1\sigma$. Grey dashed lines in bottom plot indicate $\pm 10\%$ the average continuum flux. 
    }
    \label{fig:as209_real}
\end{figure*}

AS~209 is the faintest source in HCO$^+$ $1-0$ of the five disks (Figure \ref{fig:as209_real}).
Previous observations have revealed a cold molecular cloud in the foreground of AS~209, and the cloud absorbs cold emission from AS~209 at select velocities in CO gas \citep{oberg2011,huang2016,favre2019} and {in HCO$^+$}  {gas} \citep{maps3}. 
In this source, HCO$^+$ $1-0$ is considered to be tentatively variable over the MAPS observations. 

\section{Results}\label{sec:results}

\subsection{General Behavior}

A variable source is defined as any source with at least one observation where there is a $\ge 3 \sigma$ change in  disk integrated $F_{\rm HCO^+}/F_{\rm cont}$. 
A tentatively variable source is defined as a source with at least one observation day that has variability in the HCO$^+$ J$=1-0$ intensity spectrum. Spectral variability is defined on a case by case basis in the following sections.
In this work, all changes in spectral shape are considered tentative due to (relatively) low signal to noise ratios (S/N).
A tentatively variable source does not display variations in the disk integrated $F_{\rm HCO^+}$, $F_{\rm cont}$, and $F_{\rm HCO^+}/F_{\rm cont}$, i.e. all integrated flux values are within $3 \sigma$ of each other. All tentative variability occurs on a channel by channel basis.

None of the five disks were found to be variable on a disk-integrated basis in HCO$^+$ J$=1-0$ emission during the MAPS observations, as is clear in the moment zero maps (Figure \ref{fig:momentzero}) which are constant within the error bars across nearly all epochs. 
On 03/Sep/2019, the disk integrated HCO$^+$ line flux observed in the AS~209 disk has more of an oval shape compared to the other epochs, which are more circular. This morphological difference appears to be a result of a small excess of blue shifted emission  on 03/Sep/2019 compared to the other days (discussed further in Section \ref{sec:AS209}).
On 21/Aug/2019, HCO$^+$ emission in IM~Lup is brighter than the other epochs. However, our analysis suggests the brighter line flux is likely caused by differences in baseline coverage (discussed further in Section \ref{sec:IMLup}).

While the disk integrated flux remained constant within the error bars for all disks, we report tentative variability in three of the five disks: AS~209, GM~Aur, and HD~163296. 
Each of these disks feature some form of variability or shift in the HCO$^+$ spectral shape, as further discussed below (Sections \ref{sec:AS209}, \ref{sec:GMAur},  and \ref{sec:HD163296}). 
MWC~480 also displays a shift in the HCO$^+$ spectral shape, but this shift is at the level of the noise (Section \ref{sec:MWC480}). 
IM~Lup has a brighter peak at {$6.8$ km~s$^{-1}$} than other observation days, but the higher flux is most likely caused by baseline coverage (Section \ref{sec:IMLup}). 
{Noise on a channel by channel basis is further discussed in Appendix \ref{sec:errors}. 
}

As discussed in Section~\ref{sec:baselines} and Appendix~ \ref{sec:faux}, we investigate if the differences in baseline coverage impact the measured flux. 
{A ``known flux'' model was created using {\ttfamily vis\_sample}.
The analysis} suggests that the tentative HCO$^+$ spectral variability in AS~209, GM~Aur, and HD~163296 was unlikely to be introduced by differences in baseline coverage at the level of the noise in the data. However, IM~Lup and MWC~480 appear to have been impacted by these effects.

\subsection{Continuum}\label{sec:cont}

There are no significant and distinguishable differences in the continuum images taken as a part of the MAPS observation set (Figure \ref{fig:continuum}).
For each disk, disk integrated $F_{\rm cont}$ values (taking into account RMS uncertainty) were within $10~\sim 20\%$ of each other. 
This magnitude of variation is consistent with the acceptable level of flux calibrator uncertainty reported for ALMA \citep{francis2020, ALMAhandbook}.
Therefore, all observations of continuum emission are constant within error for all disks.

While a clear ($> 3 \sigma$) change in integrated flux is not seen, the spectral peak at {$6.1$ km~s$^{-1}$} relative to the source velocity is $\sim 62$\% higher on 26/Oct/2018 compared to the other observation days. Additionally, the emission peak shifts towards the source velocity by {$\sim 5.1$} km~s$^{-1}$, or by one channel, on 26/Oct/2018. Notably, October 2018 has the best short baseline coverage, and the increase in emission and peak shift could be attributed to  {higher sensitivity  to extended emission }on this date. 

There is also some variability in the wings, notably on 3/Sep/2019 where a higher flux is is seen at {$12.1$} and {$1.6$ km~s$^{-1}$}.
Emission at {$1.6$ km~s$^{-1}$} is significant, since this is the only day that blue shifted emission is detected for any of the observed dates. While it is difficult to say with certainty, there is a slight elongation in the moment zero maps (Figure \ref{fig:momentzero}) on this day compared to the other dates. 
The channel maps showed that the excess emission was centrally peaked.

Changes in the spectrum are considered tentative, and a higher S/N is required to confirm if the general variable behavior of this source is indeed real or a noise artifact.

\begin{figure*}
    \centering
    \includegraphics[scale=0.5]{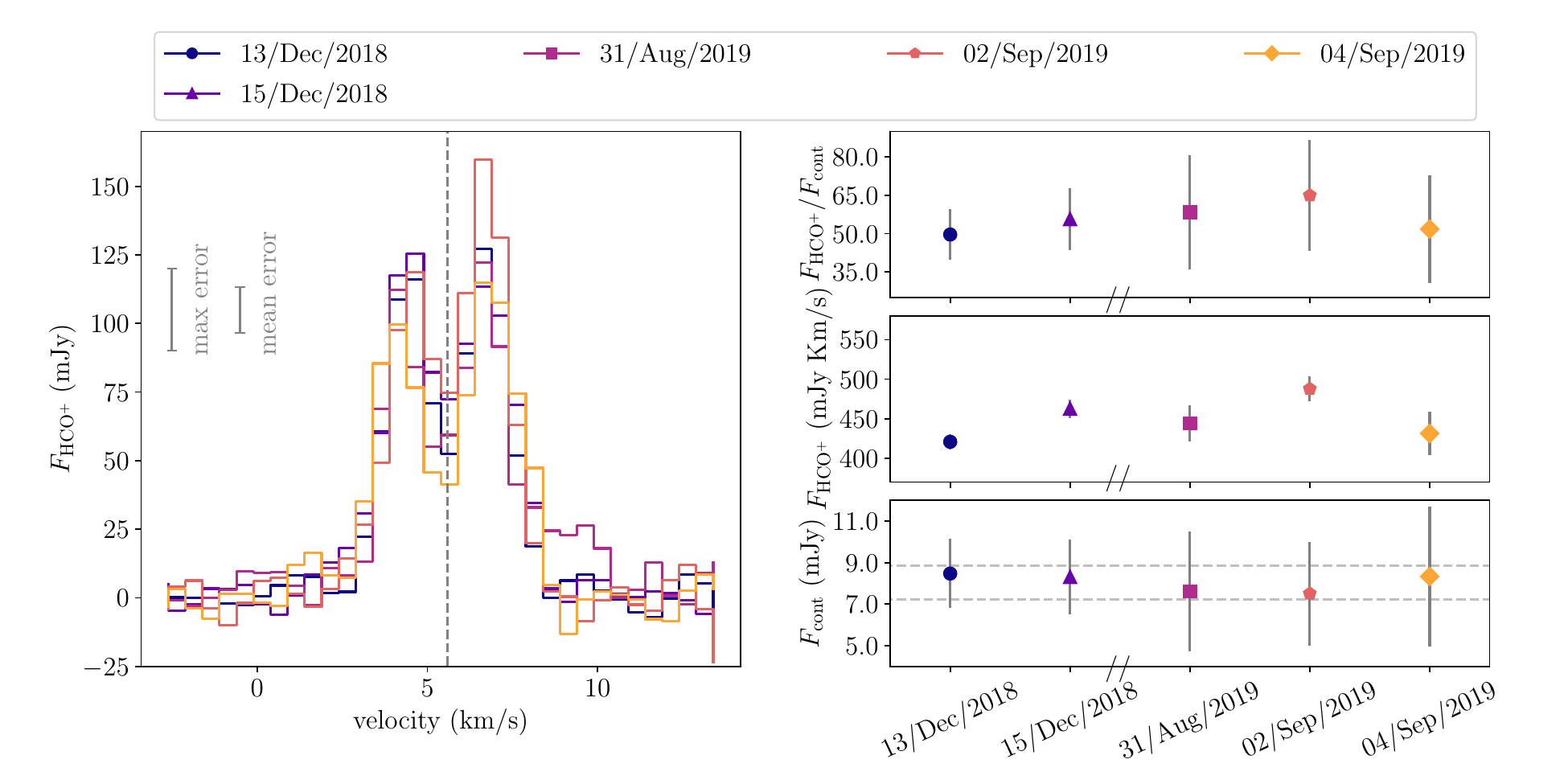}
    \caption{Same figure description as Figure \ref{fig:as209_real}, but for GM~Aur. 
    }
    \label{fig:gmaur_real}
\end{figure*}

\subsection{GM~Aur}\label{sec:GMAur}

HCO$^+$ $1-0$ is considered tentatively variable in GM~Aur due to slight variations on 31/Aug/2019 (Figure \ref{fig:gmaur_real}). On this day, peak F$_{\rm HCO^+}$ at {$6.6$ km~s$^{-1}$} is $\sim 25$\% higher than the other epochs. The peak at {$4.6$ km~s$^{-1}$} remains constant across all observations. 
On 2/Sep/2019 a shoulder is visible at {$9.6$ km~s$^{-1}$} that is not seen on the other observation days.

Notably, both occurrences of variability are asymmetric, where channels within the red shifted portion of the spectrum are enhanced while the blue-shifted side of the spectrum remains constant. 
The complex substructure observed in GM~Aur may be the force behind asymmetric variations.
As a flare propagates through the disk light could be asymmetrically scattered due to spiral arms, which were
seen in $^{12}$CO gas emission \citep{huang2021}. However, spiral arms were not seen in HCO$^+$ emission, which indicates that the substructure of GM~Aur is complex. Without further modeling it is uncertain how exactly the spiral arms would influence flare propagation or chemical changes.

 \begin{figure*}
    \centering
    \includegraphics[scale=0.5]{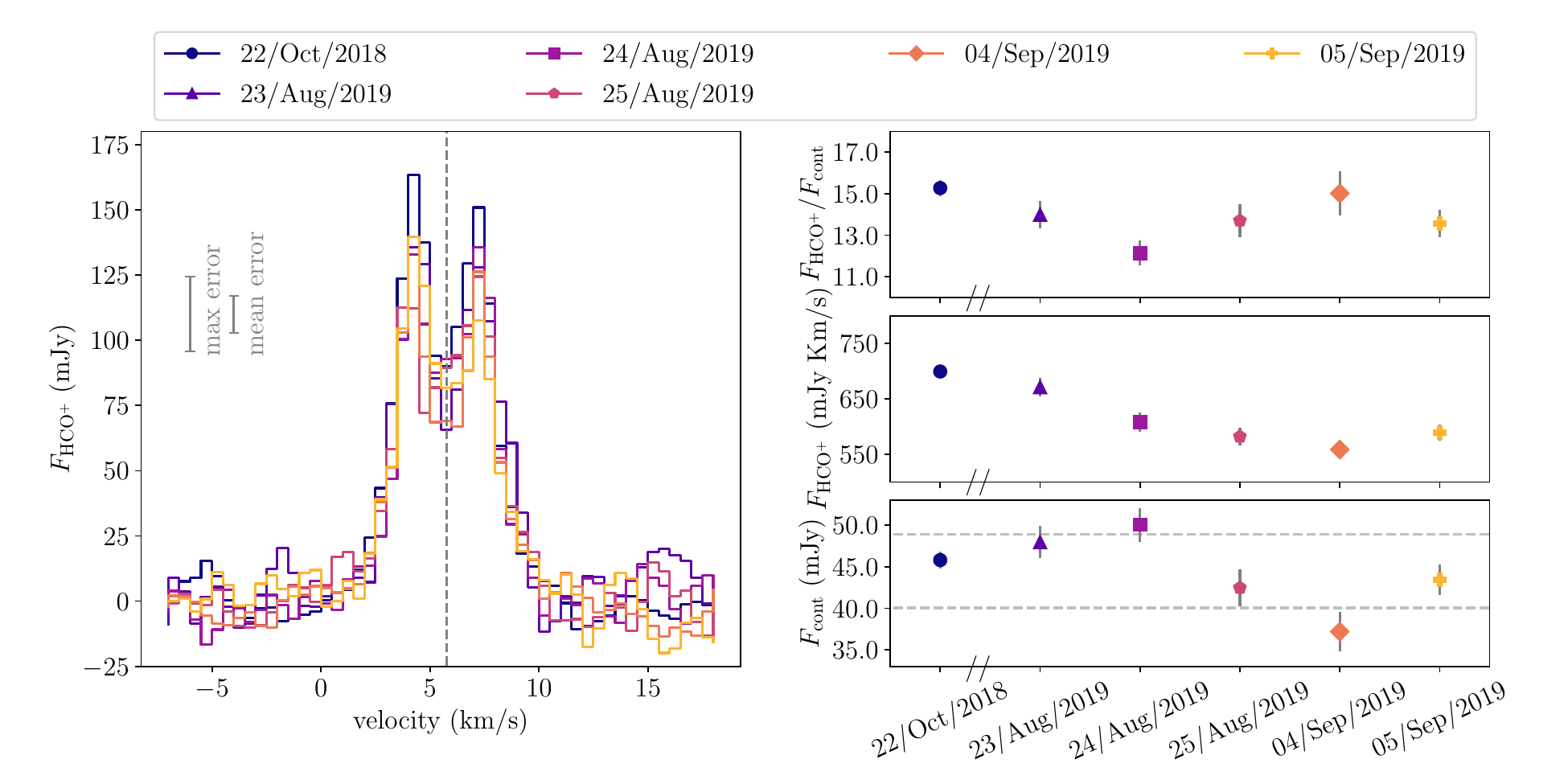}
    \caption{Same figure description as Figure \ref{fig:as209_real}, but for HD~163296. 
    }
    \label{fig:hd163296_real}
\end{figure*}

\subsection{HD~163296}\label{sec:HD163296}

HD~163296 is considered tentatively variable in HCO$^+$ $1-0$ emission (Figure \ref{fig:hd163296_real}). 
On 22/Oct/2018 (the first epoch) the double peaks are symmetrically $\sim15$\% higher than the following observations.
This change is small, symmetric, and
only occurs on one day, so calibration or baseline effects could have introduced the variation.
While steps were taken to minimize variations introduced by baseline coverage (Section \ref{sec:baselines}), the October 2018 observation has more complete low baseline coverage than the later observations. 
{While steps were taken to minimize variations introduced by baseline coverage, this epoch was the most impacted by flux loss.
There was a $42\%$ loss of flux on this day when small baselines were removed, which indicates that the October 2018 observation was strongly impacted by visibility effects. This sensitivity indicates that the enhancement could be caused by unaccounted visibilities. Even though our investigation of flux recovery did not suggest this to be true (Figure \ref{fig:fauxdata}c), the possibility is not ruled out. 
}

\subsection{IM~Lup}\label{sec:IMLup}

\begin{figure*}
    \centering
    \includegraphics[scale=0.5]{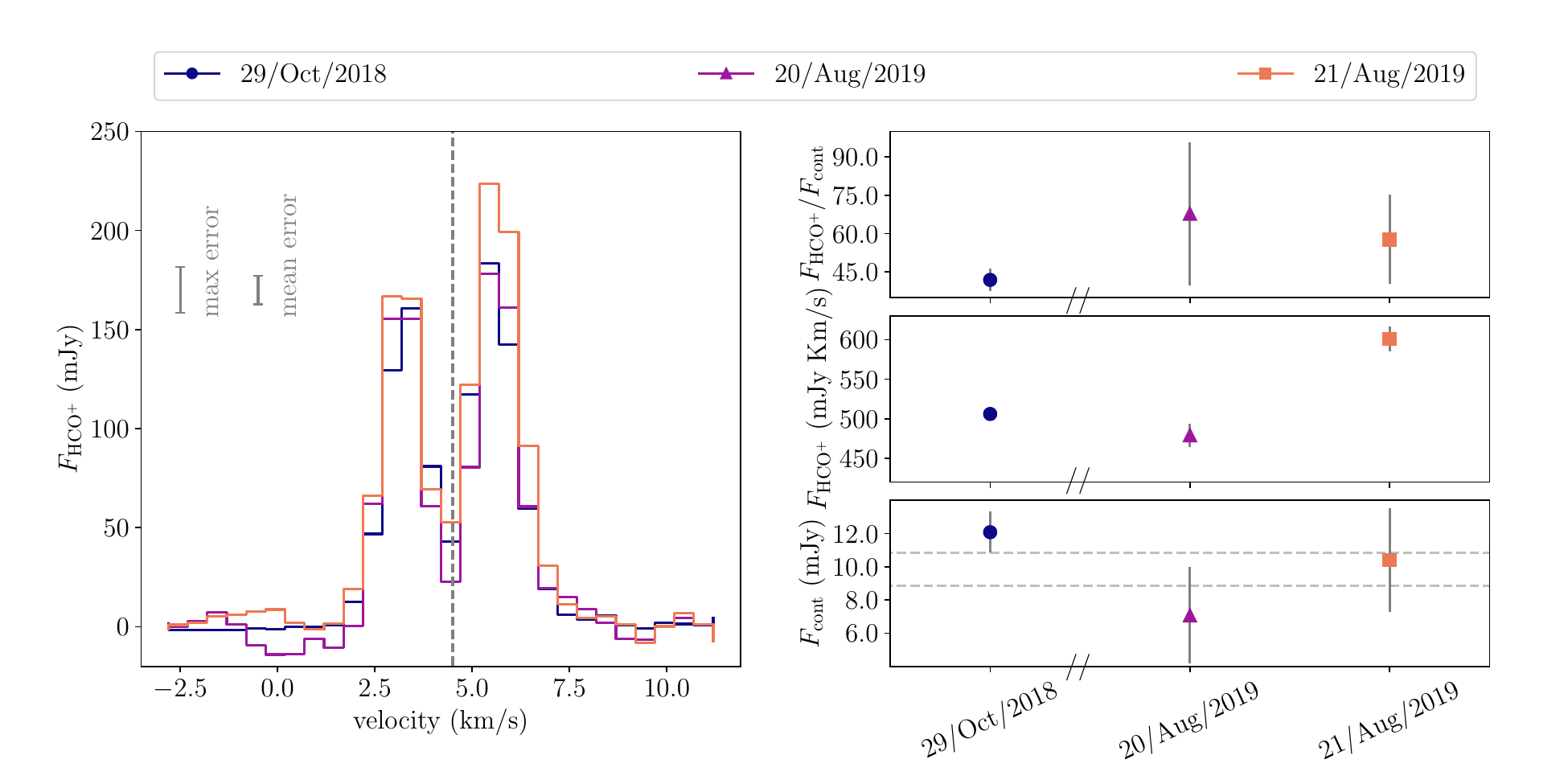}
    \caption{Same figure description as Figure \ref{fig:as209_real}, but for IM~Lup. 
    }
    \label{fig:imlup_real}
\end{figure*}

Though significant variability was seen in H$^{13}$CO$^+$ $J=3-2$ in previous observations  \citep{cleeves2017}, the HCO$^+$ $1-0$ flux is observed to be relatively constant in this data set (Figure \ref{fig:imlup_real}). 
The spectral peak at {$6.8$ km~s$^{-1}$} is $\sim26$\% higher on 21/Aug/2019 compared to other observations, which is also seen in the moment zero maps (Figure \ref{fig:momentzero}). 
However, the fluctuation may be due to baseline coverage. Flux retrieval on a model of known flux (using {\ttfamily vis\_sample}) shows the same spectral variation, suggesting that the increase in emission is introduced by differences in \textit{uv}-coverage (Figure \ref{fig:fauxdata}d). 

{IM~Lup has slightly better low baseline coverage than the other four sources, with $90\%$ of baselines being $\geq 280$m for IM~Lup and $\geq 310$m for all other sources.}  
{Additionally, IM~Lup is the most extended source \citep[gas emission up to $\sim 750$ au;][]{panic2009,avenhaus2018} in the sample, thus the large scale emission area is more sensitive to shorter baselines.}

\subsection{MWC~480}\label{sec:MWC480}

\begin{figure*}
    \centering
    \includegraphics[scale=0.5]{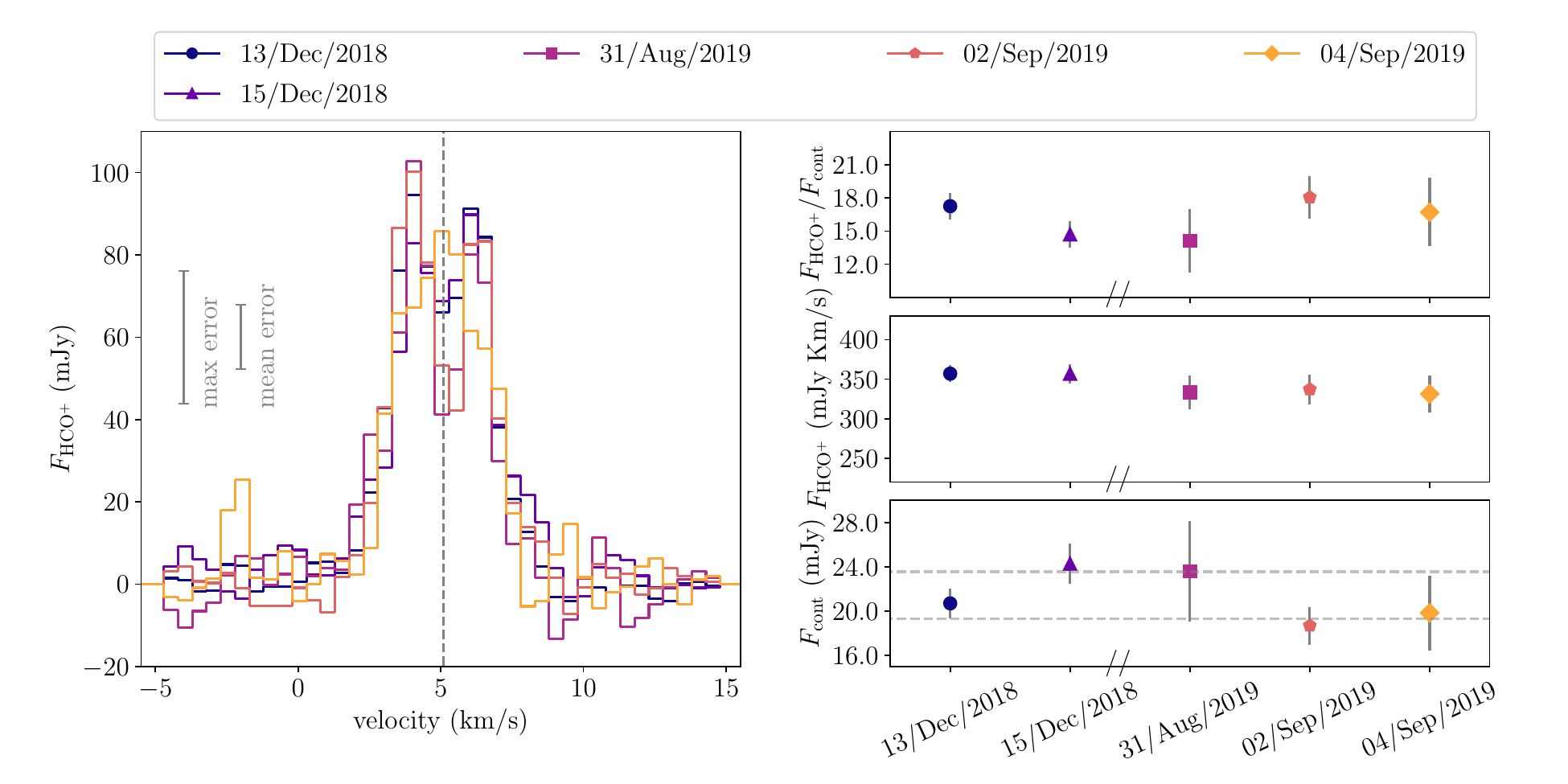}
    \caption{Same figure description as Figure \ref{fig:as209_real}, but for MWC~480. 
    }
    \label{fig:mwc480_real}
\end{figure*}

The disk integrated HCO$^+$ $1-0$ line flux in MWC~480 is effectively constant during the MAPS observations (Figure \ref{fig:mwc480_real}). 
The line shape is a symmetric double horned peak and the emission remains constant in flux and shape on all observation days except 4/Sep/2019. On this day, the spectral shape becomes symmetrically singly peaked at the source velocity ({$5.1$ km~s$^{-1}$}).  The data from this day are noisier with a lower S/N {(see Appendix \ref{sec:errors})}, therefore the single peak profile may be due to noise.

There is also a small blue-shifted peak of emission at {$-2.4$ km~s$^{-1}$} only seen on 4/Sep/2019. The emission is present across two channels and has a peak significance of $\sim 5 \sigma$ in disk-integrated line flux. Interestingly, a blue-shifted peak was also seen in earlier observations of HCO$^+$ $3-2$ as a part of the Disk Imaging Survey of Chemistry with SMA (DISCS) survey \citep{oberg2010}. At this time the origin of this tentative enhancement in emission at {$-2.4$ km~s$^{-1}$} is unclear, although it could be due to a jet \citep{grady2010} or other similar phenomena.

\section{Discussion}\label{sec:discussion}

This work was motivated by models \citep{waggoner2019,waggoner2022} and observations \citep{cleeves2017} that suggest flare driven variable X-ray ionization rates result in variable gas-phase cation abundances.
In each of the five MAPS disks, some degree of spectral variation was seen in their HCO$^+$ $1-0$ emission. However, generally the variations were small ($\lesssim 3 \sigma$), and occurred within discrete parts of the spectrum rather than an overall increase in emission.

\subsection{Possible Sources of Spectral Variability}

The two most likely scenarios that could explain spectral variations seen in this work are as follows. 
First, observed changes in flux could be  astrophysical, i.e., real variations in HCO$^+$ emission. 
Second, the changes in measured line emission could be non-astrophysical and instead introduced as a result of the observing and/or imaging process. 
In this section, we discuss the possibility of both scenarios, and the observational consequences of each scenario. 

\subsubsection{Possible Astrophysical Origins of Variability}

{In this discussion we classify possible variability sources, but it should be noted that `types' of variability may not (and are likely not) exclusive to one another. In reality variability in astronomical sources is incredibly dynamic, where bursts, flares, and other similar phenomena occur across the electromagnetic spectrum and can even include cosmic rays. 
For example, accretion outbursts increase both UV ionization rates and temperature \citep{molyarova2018}, and X-ray flares are often associated with coronal mass ejections \citep{benz2010}. Increased ionization rates driven by radiation are the most likely sources of the spectral changes we report. 
}

{
X-ray flares are one possible source of variability in the HCO$^+$ spectra. 
\citet{waggoner2019, waggoner2022} indicate that changes in abundance scale directly with flare strength, 
where larger flares result in a higher and longer increase in HCO$^+$ abundance than smaller flares.
Relatively large X-ray flares (i.e., flares that increase the baseline X-ray luminosity by tens or hundreds) can increase the abundance of HCO$^+$ by factors of $2$ or more.
Relatively small flares, also known as nano- or micro-flares \citep[i.e. flares that increase baseline X-ray luminosity by a factors of a few][]{pearce1988,feldman1997}, result in a relatively small HCO$^+$ enhancement.
Radiative transfer models are required to know precisely how flares of varying strengths impact the HCO$^+$ flux, 
so for the purpose of this discussion we assume that flux scales directly with HCO$^+$ abundance.}

{According to \citet{waggoner2022}, if a relatively large X-ray flare had occurred a clear and distinguishable change would have been seen in the HCO$^+$ spectra. Previous observations of H$^{13}$CO$^+$ in the IM~Lup protoplanetary disk support this, where the H$^{13}$CO$^+$ flux roughly doubled \citep[$28\sigma$][]{cleeves2017} in one observation compared to two others. 
Since all instances of (tentative) variation seen in this work are relatively small ($<3 \sigma$) a strong flare is unlikely to be the source of variation. However, a relatively small flare, such as a nano- or micro-flare, would drive relatively small increases in the HCO$^+$ spectrum. Therefore, if the spectral changes reported in this work were caused by an X-ray flare, they are most likely to have been caused by a relatively small flare.}

T Tauri stars, such as AS~209, GM~Aur, and IM~Lup are known to be X-ray bright and variable \citep{gudel2003}, thus suggesting that, if the fluctuations are true changes in HCO$^+$ emission, they are most likely to be caused by X-rays.
Herbig stars, such as MWC~480 and HD~163296 have bright photospheric UV emission, thus the chemistry of these disks may be more sensitive to the stellar UV radiation field, and any associated variations \citep[such as accretion or jet variability, ][]{mendigut2013, francis2020, rich2020}.
Unfortunately, with low degrees of observed variability and without concurrent X-ray or UV observations we are unable to directly confirm the {origin of the} observed spectral variations. 

Variations in the observed line emission could also be caused by changes in disk temperature instead of a change in molecular abundance. However, we find that the disk temperature - at least that in the mid-plane - remains constant to within the flux and RMS error, as indicated by the continuum measurements. Additionally, if there was a change in disk temperature the spectra would be universally and symmetrically scaled
\citep[assuming the flare uniformally impacts the disk, e.g. ][]{favata2005}
and the majority of variations in this work were found to occur asymmetrically on a channel by channel basis. 
Assuming that the HCO$^+$ $1-0$ gas and continuum trace the same temperature regime, then spectral variations would most likely be due to changes in HCO$^+$ abundance, rather than temperature. 

\subsubsection{Possible non-Astrophysical Origins of Variability}

Alternatively, the observed changes can be a product of how the disks were observed. 
Some examples include the following. 
\textit{a)}
By splitting out data on an day-to-day basis, the data is naturally noisier than the final time integrated image. Lower signal to noise, as is the case in the MWC~480 spectra, makes assessing low level variability challenging.

\textit{b)}
We minimized the effect of baseline coverage as much as possible (see Sections \ref{sec:baselines} and Appendix~\ref{sec:faux}); however,  
baseline effects cannot be ruled out for HD~163296, the shift in AS~209 emission peak, or in IM~Lup, since these instances of variability occurred when ALMA was in a compact configuration. 
Our flux recovery tests indicate that the variability in IM~Lup was likely caused by spatial filtering, but HD~163296 and AS~209 variability likely has a different source.

\textit{c)}  
Flux calibrators are ideally considered to be constant on observationally relevant time scales, but this is not always the case. {In reality, minor changes in flux calibrator luminosity do occur. If a flux calibrator varied in brightness, a symmetric and uniform change would be introduced in the HCO$^+$ spectra and continuum. 
The variations reported here are asymmetric and non-uniform (except HD~163296).}
Therefore, we find that the changes in flux calibrator are an unlikely source of variability. 
For a complete list of flux calibrators used in this work, see \citet{maps1}.

\subsection{Tentative First Detection of Spatial Variability in Disks}

AS~209 and GM~Aur are the first protoplanetary disks with potential spatial variability, as indicated by a shift in the spectral peak and variations in the line wings of the spectra. 
The physical origin of the peak shifts and fluctuations in spectral wings (if real) cannot be confirmed due to the timing and limit of observation windows, but a reasonable explanation is that the shift was caused by an X-ray flare propagating through the disk. This discussion assumes an enhancement of HCO$^+$ immediately (within minutes) traces the flare, as indicated by \citet{waggoner2022}.

Light produced by large X-ray flares (i.e., a flare that increases the characteristic luminosity by $>10 \times$) 
will symmetrically propagate 
through the disk
\citep[due to the large size of the X-ray emitting region compared to the physical size of the star, e.g.][]{favata2005}. 
 If a disk were to be observed immediately after a flare occurs, i.e., after the flare hits the inner disk but has not yet reached the outer disk (hours after the flare), then an increase in inner disk HCO$^+$ emission would be seen.
 This effect would primarily be seen in the high velocity spectral wings, as seen on 03/Sep/2019 for AS~209 and on 31/Aug/2019 for GM~Aur.

If the disk were to be re-observed after the flare has propagated through the entire disk (over a timescale longer than the light crossing time, $\sim$days), then a uniform increase in the spectrum would be seen. 
A uniform enhancement occurred for HCO$^+$ in HD~163296 in this work and H$^{13}$CO$^+$ in IM~Lup in \citet{cleeves2017}, suggesting that these observations occurred several days after the flare occurred.

After the flare has propagated through the entire disk, inner disk HCO$^+$ molecules will rapidly dissociate via electron recombination to H and CO, while the outer disk HCO$^+$ abundance remains enhanced for a longer period due to a lower level of free electrons. If HCO$^+$ rotational emission were observed during this time, the spectrum closer to the source velocity, including the peaks, would be enhanced while the wings would be quiescent. The spectra observed on 26/Oct/2018 for AS~209 and on 02/Sep/2019 for GM~Aur are consistent with this interpretation.
How flares change the chemistry as a function of radius and time is described further in  \citet{waggoner2022}, who report the same phenomena in models of HCO$^+$ column density throughout a flaring event \citep[see Figure 7 in ][]{waggoner2022}.

Large X-ray flares originating within extended stellar coronae are expected  to uniformly illuminate the disk, changing the HCO$^+$ abundance across the entire azimuthal range. But this is not consistent with the type of variations seen in AS~209 and GM~Aur. 
Since only half of the AS~209 spectrum is visible due to cloud obstruction, we are unable to determine if the spectral changes were indeed spectrally (and spatially) symmetric. Future observations targeting warmer HCO$^+$ lines, which are expected to be less impacted by foreground absorption, are necessary to 
explore symmetric variations, as shown by \citet{oberg2011} and \citet{huang2017}. 
GM~Aur has a known $^{12}$CO spiral structure \citep{huang2021}.
When a flare occurs in this system, X-ray light would still uniformly propagate outward. Instead of uniformly impacting the disk and enhancing HCO$^+$, the flare could first impact the closest arm, thus causing an asymmetric increase in the HCO$^+$ spectrum. 
This cannot be confirmed without additional chemical modeling including 3D structure; however, both cases where $F_{\rm HCO^+}$ increased in GM~Aur only occurred in the red-shifted portion of the spectrum.

\subsection{Tentative First Detection of Variability in a Herbig System}

We report the first tentative detection of HCO$^+$ variability in a Herbig protoplanetary disk system, HD~163296. 
Unfortunately, the peak HCO$^+$ emission occurred during the first observation window, and the following observation windows did not occur until a year later. 
If HD~163296's enhancement was real, the duration and behavior are unknown since it was not re-observed for 9 months. Thus it is challenging to determine the exact cause of HCO$^+$ variability in this source. While the source of variability could be an X-ray flare, it is also possible that a fluctuation of UV radiation could have led to the higher HCO$^+$ emission observed in October 2018.

Fluctuations in UV flux may be a more likely origin than X-rays in HD~163296, as Herbig stars are more massive and hotter than solar mass T Tauri stars. Stellar spectrum for Herbig star emission peaks in the UV range, resulting in much lower relative contributions from X-ray emission than T Tauri stars. Therefore, Herbig disk ionization may instead be driven by a variable UV flux. 
Since HCO$^+$ is sensitive to disk ionization rates, which is connected to both UV and X-ray flux \citep[e.g.][]{seifert2021}, it is impossible to determine the exact origin of variability in HD~163296 without further multi-wavelength monitoring observations.

\subsection{Comparison to X-ray Flare Models}

How does the degree of HCO$^+$ variability compare with that predicted by chemical models including X-ray flares? 
Of the five sources observed by MAPS, none showed strong variability at the level previously seen in IM~Lup as reported in \citet{cleeves2017}. X-ray flares are known to occur on a regular basis \citep[varying by factors of $4$ to $10$ about once a week for T Tauri stars,][]{wolk2005}. Stronger flares are rarer; they occur only every few months (varying by factors greater than $10$.)

We can compare this lack of strong chemical variability with disk models reported in \citet{waggoner2022}. Based on the typical uncertainty in the measured disk integrated flux, a minimum change of \textit{at least} $100\%$ in disk integrated $F_{\rm HCO^+}$ is required to confidently detect enhancement ($>3\sigma$). A smaller change may be detectable in a disk with a higher signal to noise ratio, but this discussion uses the typical flux and uncertainties in this work.
Flare models predict a $2.2\%$ chance of observing the disk-integrated HCO$^+$ enhanced by a factor of at least $100\%$ when observed one time on any given day. 
Each of the MAPS disks was observed between 3 and 6 times, so we can estimate the probability that they are observed to be enhanced during at least one of these observations. 

In total, 24 separate observations were executed.
If this entire data set is considered, then there is a $41\%$ chance of observing clear and distinguishable variability. 
If Herbig (observed 11 times) and T Tauri systems (observed 13 times) are considered separately, then there is only $22\%$ and $25\%$ chance of observing enhancement, respectively.
For the HD~163296 disk (observed six times), there was a $12.5\%$ chance of observing HCO$^+$ to be enhanced by $100\%$ or greater at least one time. For the disks observed five times, AS~209, GM~Aur, or MWC~480, there was a $10.5\%$ chance of observing them with elevated HCO$^+$. Finally, IM~Lup was only observed three times, thus $6.5\%$ chance of observing variability in IM~Lup.

Note that the theoretical probabilities discussed here are likely an overestimate of the chance of detecting a flare, since many of the MAPS observations occurred within several days with each other, while the modeled statistics are derived from a 500 year window and assume  observations are independent. 
This discussion also assumes HCO$^+$ emission increases linearly with an increase in abundance, which is only true if the line is optically thin and the emitting conditions are constant. 
Additionally, these statistics are true for X-ray flares produced by T Tauri stars, and the probability of observing X-ray or UV driven variability in a Herbig star is likely different.

\section{Conclusions}\label{sec:conclusions}

Fluctuations in ionization, such as an X-ray flaring event, have been shown to temporarily increase the abundance of gas-phase cations, which results in time variable emission.
In this work we searched for variability in HCO$^+$ J=$1-0$ in the HD~163296, AS~209, GM~Aur, MWC~480, and IM~Lup protoplanetary disks using data from the MAPS ALMA Large Program. 
While disk integrated $F_{\rm HCO^+}$ remained constant within the uncertainty across all observations, low level spectral variability was seen in all five disks.

\begin{itemize}
    \item AS~209: 
         On 26/Oct/2018 the HCO$^+$ spectral peak is increased and shifted blueward toward the source velocity.
         While this particular shift could be impacted by baseline coverage, there is also variation in the spectral wings and an increase in blue shifted emission is also seen on 03/Sep/2019. 
    \item GM~Aur:
          There is an asymmetric increase in peak emission on 02/Sep/2019, and 
          enhancement in the red-shifted wing is also seen on 31/Aug/2019.
          The asymmetric spectral shifts may be due to disk structural asymmetry, as GM~Aur has spiral arms in $^{12}$CO gas emission. 
    \item HD~163296: 
          The spectral peaks are slightly enhanced on 22/Oct/2018 compared to later observations. While this could be due to an increase in ionization rates, this fluctuation could also be due to varying baseline coverage. 
    \item IM~Lup: 
        There is an asymmetric increase in emission for the red shifted peak on 21/Aug/2019, but this same phenomena occurs in data simulated using {\ttfamily vis\_sample}. Therefore, the spectral variation in IM~Lup is most likely introduced by differences in baseline coverage between observations.
    \item MWC~480: 
          There is a change in spectral shape on 04/Sep/2019, where the spectrum has a single peak compared to two peaks on the other observation days. However, the change in shape can be attributed to higher levels of noise and is not conclusively astrophysical. 
\end{itemize}

Our {observations 
are consistent} with the theory that a temporary increase in high energy stellar emission
can {drive changes in the abundance and flux of gas-phase cations in planet-forming disks}
\citep{waggoner2022}.

Additionally, we show that substantial serendipitous fluctuations in disk-integrated flux as was seen in IM~Lup \citep{cleeves2017} are rare. In order to confidently detect flare driven chemistry and constrain spectral variation, a dedicated time domain campaign, optimizing high S/N, is required. 
A designated time domain observing program is necessary to measure variable gas-phase cation emission. Such a program will be scientifically useful, as it could result in measurements of electron abundances and information about magnetic fields, magnetorotational instability (MRI), and disk accretion \citep[e.g. ][]{balbus1991,glassgold1997,ilgner2006b}.

\acknowledgments
{We thank the anonymous referees for their insightful comments on this work.}
We thank the entire MAPS collaboration for all the hard work that went into data collection and reduction. 
This paper makes use of the following ALMA data: ADS/JAO.ALMA\#2018.1.01055.L.
ALMA is a partnership of ESO (representing its member states), NSF (USA) and NINS (Japan), together
with NRC (Canada), MOST and ASIAA (Taiwan), and
KASI (Republic of Korea), in cooperation with the Republic of Chile. The Joint ALMA Observatory is operated by ESO, AUI/NRAO and NAOJ. The National
Radio Astronomy Observatory is a facility of the National Science Foundation operated under cooperative
agreement by Associated Universities, Inc.
A.R.W. acknowledges support from the
Virginia Space Grant Consortium and the National Science
Foundation Graduate Research Fellowship Program under grant
No. 1842490. Any opinions, findings, and conclusions or
recommendations expressed in this material are those of the
author(s)and do not necessarily reflect the views of the National
Science Foundation.
L.I.C. acknowledges support from the David and Lucille Packard Foundation, Research Corporation for Science Advancement, NASA ATP 80NSSC20K0529, and NSF grant no. AST-2205698.
K.I.\"O. acknowledges support from the Simons Foundation (SCOL \#321183), an award from the Simons Foundation (\#321183FY19), and an NSF AAG Grant (\#1907653).

\appendix

\section{Visibility Effects on HCO$^+$ Emission}\label{sec:faux}

To ensure that any detected variability was not artificially introduced by varying telescope configurations, the python function {\ttfamily vis-sample} was used to generate `mock' data (as described in Section \ref{sec:baselines}). These data are produced using the exact same {\ttfamily CLEAN}ing method as the real data. Therefore, the mock data reveal any variability that would have been introduced by varying baseline coverage from different telescope configurations.

\begin{figure}[h!]
    \centering
    \includegraphics[scale=0.28]{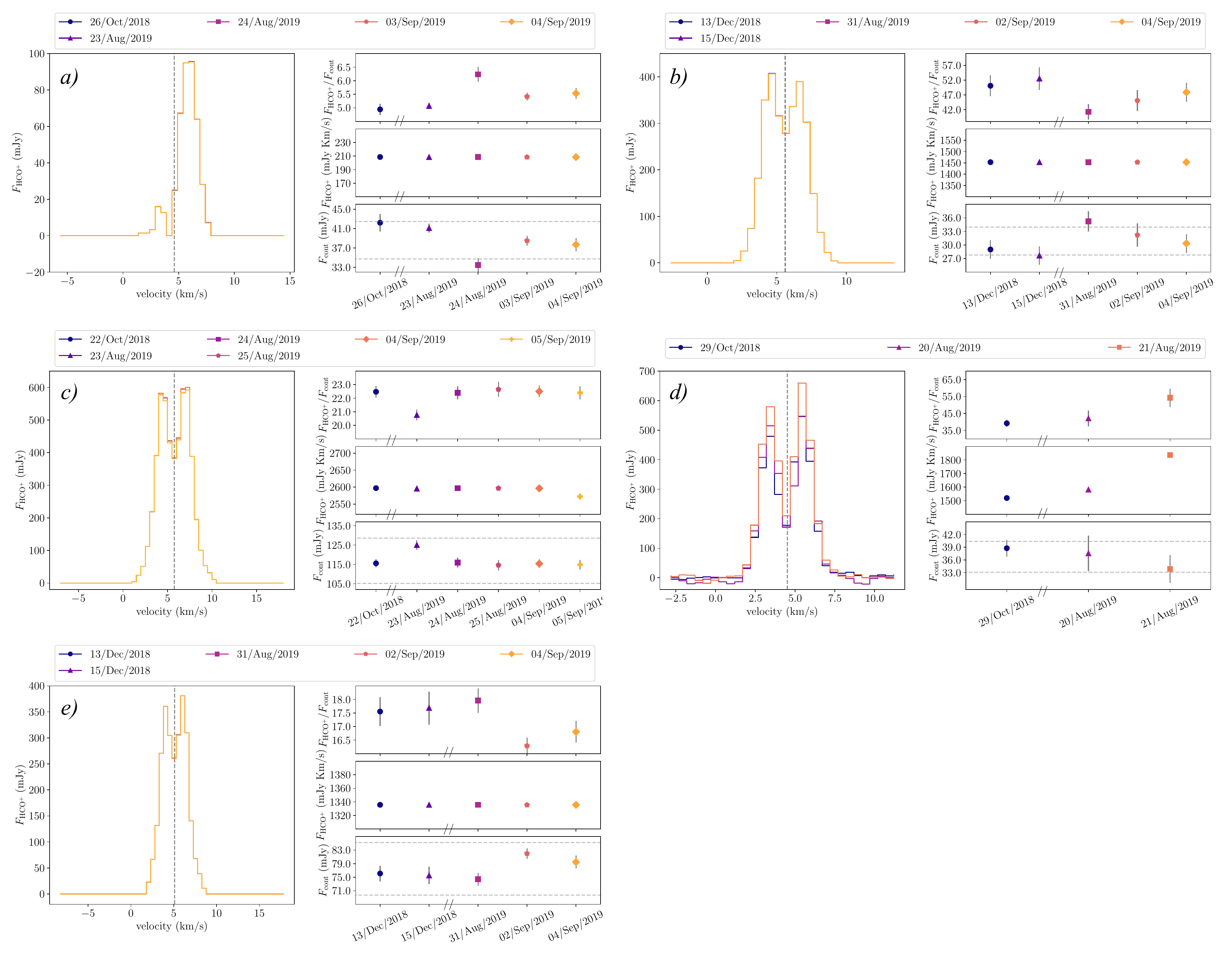}
    \caption{Synthetic spectra (``mock data'') produced using the {\ttfamily vis-sample} routine on the time-integrated HCO$^+$ $1-0$ cubes. These models reveal any spectral or flux fluctuations that would have been introduced by varying baseline coverage for each source's unique spatial emission structure. Each sub-figure shows the analysis for:
    \textit{a)} AS~209;
    \textit{b)} GM~Aur;
    \textit{c)} HD~163296;
    \textit{d)} IM~Lup;
    \textit{e)} MWC~480.
    For each panel, \textit{Left:} mock HCO$^+$ $1-0$ spectrum, and \textit{Right:} mock disk integrated continuum, line, and continuum normalized line emission. Note that the spectra overlap.
    }
    \label{fig:fauxdata}
\end{figure}

IM~Lup is the only system with a clear baseline effect in the mock data. $F_{\rm HCO^+}$ on 21/Aug/2019 is higher than the other observations, despite the fact that all mock observations were modeled using the same initial measurement set. This same effect is seen in the real data (Figure \ref{fig:imlup_real}), and is attributed to baseline coverage.

There are several scenarios where observations with more complete short baseline coverage (i.e.,\ those taken in October 2018) yield slightly higher $F_{\rm HCO^+}$ values (HD~163296, AS~209). 
In the disks where this effect occurs, no significant variations were seen in the mock line or continuum flux. 
There are variations in the disk integrated flux values, most notably in $F_{\rm cont}$. This variation is attributed to RMS uncertainty, since all $F_{\rm cont}$ values are with $\pm 10\%$ of each other (See Section \ref{sec:cont}). 
Changes in flux are unlikely {caused by baseline} coverage.

\section{Channel Error Bars on Spectra}\label{sec:errors}

\begin{figure}
    \centering
    \includegraphics[scale=0.5]{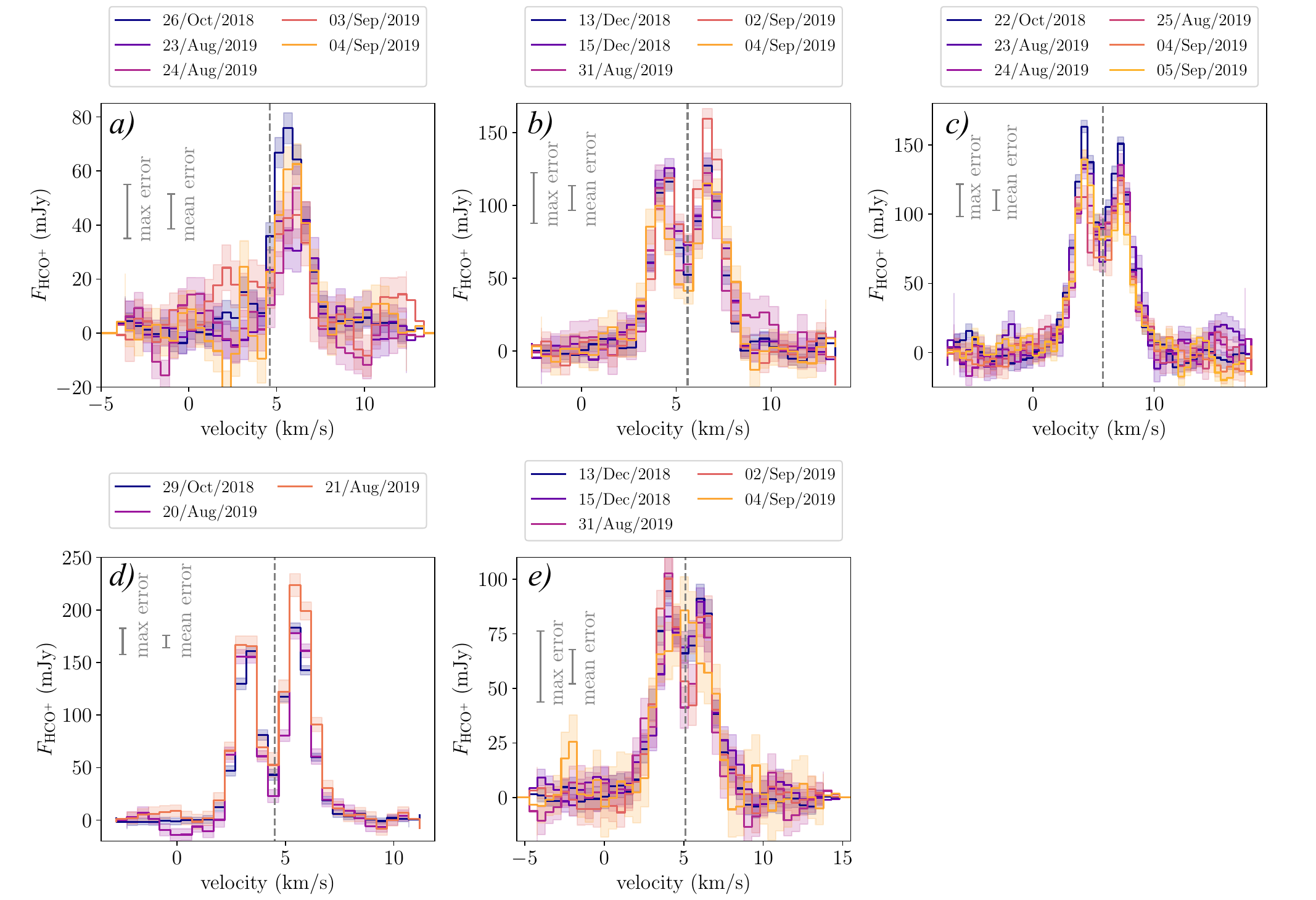}
    \caption{{HCO$^+$ J$=1-0$ spectra for each of the MAPS disks with error bars for each channel. The grey error bars represent the max and mean error bar (standard deviation) for all channels.  
        \textit{a)} AS~209;
    \textit{b)} GM~Aur;
    \textit{c)} HD~163296;
    \textit{d)} IM~Lup;
    \textit{e)} MWC~480.
    \label{fig:errorbars}}}
\end{figure}

{
All spectral changes reported in this work are considered tentative, since no change is $>3\sigma$. 
However, there are a number of spectral variations greater than the level of noise, as shown in Figure \ref{fig:errorbars} and defined as follows. 
\begin{itemize}
    \item AS~209: The peak at 6.1 km~s$^{-1}$ is enhanced and shifted toward source velocity on 26/Oct/2018. The 6.1 km~s$^{-1}$ peak appears to also be shifted and enhanced on 04/Sep/2019, but not above the level of noise, so it is not considered a change in emission in this work. 
    On most observation days little to no blue shifted emission is seen, except on 03/Sep/2019.
     \item  GM~Aur: The peak at $6.6$ km~s$^{-1}$ is enhanced on 26/Oct/2018, and a shoulder is visible at $9.6$ km~s$^{-1}$ on 2/Sep/2019. 
    \item HD~163296: Both central peaks are enhanced on 22/Oct/2018 compared to the following observations. 
    \item IM~Lup: The peak at $6.8$ km~s$^{-1}$ is enhanced on 21/Aug/2019 compared to the other two observations. 
    \item MWC~480: An unknown emission peak is seen on 04/Sep/2019 at km~s$^{-1}$ km/s and not on any other observation day. There may be a shift in spectral shape on 04/Sep/2019, but the change is below the level of noise and not considered in this work.
\end{itemize}}

{For more information on error calculations, see Section \ref{sec:fluxanderror}.}

\end{document}